\documentclass[journal]{IEEEtran}

\IEEEoverridecommandlockouts

\usepackage{graphicx}
\usepackage{amsmath}
\usepackage{amssymb}
\usepackage{verbatim}
\usepackage{cite}
\usepackage{url}
\usepackage{cases}
\usepackage{subfigure}

\begin{document}

\title{\LARGE Computationally Efficient Modulation Level Classification Based on Probability Distribution Distance Functions}
\author{Paulo Urriza, Eric Rebeiz, Przemys{\l}aw Pawe{\l}czak, and Danijela \v{C}abri{\'c}%
\thanks{The authors are with the Department of Electrical Engineering, University of California, Los Angeles, 56-125B Engineering IV Building, Los Angeles, CA 90095-1594, USA (email: \{pmurriza, rebeiz, przemek, danijela\}@ee.ucla.edu).}
}
\maketitle
\begin{abstract}
We present a novel modulation level classification (MLC) method based on probability distribution distance functions. The proposed method uses modified Kuiper and Kolmogorov-Smirnov distances to achieve low computational complexity and outperforms the state of the art methods based on cumulants and goodness-of-fit tests. We derive the theoretical performance of the proposed MLC method and verify it via simulations. The best classification accuracy, under AWGN with SNR mismatch and phase jitter, is achieved with the proposed MLC method using Kuiper distances.
\end{abstract}

\IEEEpeerreviewmaketitle
\vspace{-0.2cm}
\section{Introduction}
\label{sec:introduction}

Modulation level classification (MLC) is a process which detects the transmitter's digital modulation level from a received signal, using a priori knowledge of the modulation class and signal characteristics needed for downconversion and sampling. Among many modulation classification methods~\cite{dobre_ietcomm_2007}, a cumulant (Cm) based classification~\cite{swami_tcom_2002} is one of the most widespread for its ability to identify both the modulation class and level. However, differentiating among cumulants of the same modulation class, but with different levels, i.e. 16QAM vs. 64QAM, requires a large number of samples. A recently proposed method~\cite{wang_twc_2010} based on a goodness-of-fit (GoF) test using Kolmogorov-Smirnov (KS) statistic has been suggested as an alternative to the Cm-based level classification which require lower number of samples to achieve accurate MLC.

In this letter, we propose a novel MLC method based on distribution distance functions, namely Kuiper (K)~\cite{cirrone_tns_2004}~\cite[Sec. 3.1]{Stephens_jasa_1974} and KS distances, which is a significant simplification of methods based on GoF. We show that using a classifier based only on K-distance achieves better classification than the KS-based GoF classifier. At the same time, our method requires only $2ML$ additions in contrast to $2M(\log2M+2K)$ additions for the KS-based GoF test, where $K$ is the number of distinct modulation levels, $M$ is the sample size and $L\ll M$ is the number of test points used by our method.

\vspace{-0.2cm}

\section{Proposed MLC Method}
\label{sec:modulation_classification}

\subsection{System Model}
\label{sec:model}

Following~\cite{wang_twc_2010}, we assume a sequence of $M$ discrete, complex, i.i.d. and sampled baseband symbols, $\mathbf{s}^{(k)}\triangleq[s_1^{(k)} \cdots s_M^{(k)}]$, drawn from a modulation order $\mathcal{M}_k\in\left\{\mathcal{M}_1,\ldots,\mathcal{M}_K\right\}$, transmitted over AWGN channel, perturbed by uniformly distributed phase jitter and attenuated by an unknown factor $A>0$. Therefore, the received signal is given as $\mathbf{r}\triangleq[r_1 \cdots r_M]$, where $r_n= A e^{j\Phi_n}s_n+g_n$, $\{g_n\}_{n=1}^M\sim\mathcal{CN}\left(0,\sigma^{2}\right)$ and $\{\Phi_n\}_{n=1}^M\sim\mathcal{U}\left(-\phi,+\phi\right)$. The task of the modulation classifier is to find $\mathcal{M}_k$, from which $\mathbf{s}^{(k)}$ was drawn, given $\mathbf{r}$. Without loss of generality, we consider unit power constellations and define SNR as $\gamma\triangleq A^2/\sigma^2$.

\vspace{-0.2cm}

\subsection{Classification based on Distribution Distance Function}
\label{sec:proposed_method}

The proposed method modifies MLC technique based on GoF testing using the KS statistic~\cite{wang_twc_2010}. Since the KS statistic, which computes the minimum distance between theoretical and empirical cumulative distribution function (ECDF), requires \emph{all} CDF points, we postulate that similarly accurate classification can be obtained by evaluating this distance using a smaller set of points in the CDF.

Let $\mathbf{z}\triangleq[z_1 \cdots z_N]=f(\mathbf{r})$ where $f(\cdot)$ is the chosen feature map and $N$ is the number extracted features. Possible feature maps include $|\mathbf{r}|$ (magnitude, $N=M$) or the concatenation of $\Re\{\mathbf{r}\}$ and $\Im\{\mathbf{r}\}$ (quadrature, $N=2M$). The theoretical CDF of $z$ given $\mathcal{M}_k$ and $\gamma$, $F_0^k(z)$, is assumed to be known a priori (methods of obtaining these distributions, both empirically and theoretically, are presented in~\cite[Sec. III-A]{wang_twc_2010}). The $K$ CDFs, one for each modulation level, define a set of test points
\begin{equation}
t_{ij}^{(\epsilon)}=\arg\max_{z}D_{ij}^{(\epsilon)}(z),
\label{eq:testpoints}
\end{equation}
with the distribution distances given by
\begin{equation}
D_{ij}^{(\epsilon)}(z)=(-1)^{\epsilon}\left(F_{0}^{i}(z)-F_{0}^{j}(z)\right),
\end{equation}
for $1\leq i,j \leq K$, $i\neq j$, and $\epsilon \in \{0,1\}$, corresponding to the maximum positive and negative deviations, respectively.  Note the symmetry in the test points such that $t_{ji}^{(0)}=t_{ij}^{(1)}$. Thus, there are $L\triangleq 2\binom{K}{2}$ test points for a $K$ order classification.

The ECDF, given as
\begin{equation}
F_{N}(t)=\frac{1}{N}\sum\limits_{n=1}^{N}\mathcal{\mathbb{I}}(z_n\leq t),
\end{equation}
is evaluated at the test points to form $\mathbf{F}_N\triangleq\{F_{N}(t_{ij}^{(\epsilon)})\}$, ${1\leq i,j \leq K, i\neq j}$. Here, $\mathbb{I}(\cdot)$ equals to one if the input is true, and zero otherwise. By evaluating $F_{N}(t)$ only at the test points in (\ref{eq:testpoints}), we get
\begin{equation}
\hat{D}^{(\epsilon)}_{ij} = (-1)^{\epsilon}\left(F_{N}\left(t^{(\epsilon)}_{ij}\right)-F_{0}^j\left(t_{ij}^{(\epsilon)}\right)\right)
\label{eq:dij}
\end{equation}
which are then used to find an estimate of the maximum positive and negative deviations
\begin{equation}
\hat{D}_{j}^{(\epsilon)}=\max_{1\leq i\leq K,i\neq j} \hat{D}^{(\epsilon)}_{ij},\quad 1\leq j \leq K,
\label{eq:dj}
\end{equation}
of the ECDF to the true CDFs. The operation of finding the ECDF at the given testpoints (\ref{eq:dij}) can be implemented using a simple thresholding and counting operation and does not require samples to be sorted as in~\cite{wang_twc_2010}. The metrics in (\ref{eq:dj}) are used to find the final distribution distance metrics
\begin{equation}
\hat{D}_{j} =\max\left(\left|\hat{D}_{j}^{(0)}\right|,\left|\hat{D}_{j}^{(1)}\right|\right),\quad\hat{V}_j = \left|\hat{D}^{(0)}_j + \hat{D}^{(1)}_j\right|,
\label{eq:metrics}
\end{equation}
which are the reduced complexity versions of the KS distance (rcKS) and the K distance (rcK), respectively\footnote{Note, that other non-parametric distances used in hypothesis testing exist (see introduction in e.g.~\cite{cirrone_tns_2004}), although for brevity they are not addressed here. We note, however, that our approach is easily applied to any assumed distance metric.}. Finally, we use the metrics in (\ref{eq:metrics}) as substitutes to the true distance-based classifiers with the following rule: choose $\mathcal{M}_{\hat{k}}$ such that
\begin{equation}
\hat{k}_D= \arg \min_{1\leq j \leq K} \hat{D}_j,  \
\hat{k}_V=\arg \min_{1\leq j \leq K} \hat{V}_j.
\label{eq:classifiers_rcKs}
\end{equation}
In the remainder of the letter, we define $h_{\hat{D}}(\mathbf{F}_N)=\hat{k}_D$ and $h_{\hat{V}}(\mathbf{F}_N)=\hat{k}_V$, where $\hat{k}_D, \hat{k}_\text{V} \in \{1,\ldots,K\}$.

\vspace{-2.5mm}

\subsection{Analysis of Classification Accuracy}
\label{sec:analysis}

Let $\mathbf{t}\triangleq[t_1 \cdots t_L]$ denote the set of test points, $\{t_{ij}^{(\epsilon)}\}$, sorted in ascending order. For notational consistency, we also define the following points, $t_0\triangleq-\infty$ and $t_{L+1}\triangleq+\infty$. Given that these points are distinct, they partition $\mathbf{z}$ into $L+1$ regions. An individual sample, $z_n$, can be in region $l$, such that $t_{l-1} < z_n \leq t_{l}$, with a given probability, determined by $F_0^k(z)$.

Assuming $z_n$ are independent of each other, we can conclude that given $\mathbf{z}$, the number of samples that fall into each of the $L+1$ regions, $\mathbf{n}\triangleq [n_1 \cdots n_{L+1}]$, is jointly distributed according to a multinomial PMF given as
\begin{equation}
f(\mathbf{n} | N,\mathbf{p}) = 
\begin{cases}
\frac{N! p_1^{n_1} \cdots p_{L+1}^{n_{L+1}}}{{n_1}! \cdots {n_{L+1}}!},& \text{if $\sum\limits_{i=1}^{{L+1}}n_i=N$},\\
0,& \text{otherwise},
\end{cases}
\label{eq:multinomial}
\end{equation}
where $\mathbf{p}\triangleq [p_1 \cdots p_{L+1}]$, and $p_l$ is the probability of an individual sample being in region $l$. Given that $\mathbf{z}$ is drawn from $\mathcal{M}_k$, $p_l=F_0^k(t_l)-F_0^k(t_{l-1})$, for $0<l\leq L+1$.

Now, with particular $\mathbf{n}$, the ECDF at all the test points is
\begin{equation}
\mathbf{F}_N(\mathbf{n})\triangleq[F_N(t_1) \cdots F_N(t_{L})],\quad F_N(t_l) = {\frac{1}{N}} \sum\limits_{i=1}^l n_i.
\end{equation}
Therefore, we can analytically find the probability of classification to each of the $K$ classes as
\begin{equation}
\Pr(\hat{k} = \kappa | \mathcal{M}_k) =\!\!\!\sum\limits_{\mathbf{n}\in\mathbb{N}^{L+1}}\!\!\!\mathbb{I}(h_{\hat{V}}(\mathbf{F}_N(\mathbf{n}))=\kappa)f(\mathbf{n} | N,\mathbf{p}),
\label{eq:ProbabilityOfClassification}
\end{equation}
for the rcK classifier. A similar expression can be applied to rcKS, replacing $h_{\hat{V}}(\cdot)$ with $h_{\hat{D}}(\cdot)$ in (\ref{eq:ProbabilityOfClassification}).

\subsection{Complexity Analysis}
\label{sec:complexity}

Given that the theoretical CDFs change with SNR, we store distinct CDFs for $W$ SNR values for each modulation level (impact of the selection of $W$ on the accuracy is discussed further in Section \ref{sec:detection_samples}.) Further, we store $KW$ theoretical CDFs of length $\bar{N}$ each. For the non-reduced complexity classifiers that require sorting samples, we use a sorting algorithm whose complexity is $N\log N$. From Table~\ref{table:complexity}, we see that for $K \leq 3$ rcK/rcKS tests use less addition operations than K/KS-based methods~\cite{wang_twc_2010} and Cm-based classification~\cite{swami_tcom_2002}. For $K>3$, the rcK method is more computationally efficient when implemented in ASIC/FPGA, and is comparable to Cm in complexity when implemented on a CPU. In addition, the processing time would be shorter for an ASIC/FPGA implementation, which is an important requirement for cognitive radio applications. Furthermore, their memory requirements are also smaller since $\bar{N}$ has to be large for a smooth CDF. It is worth mentioning that the authors in~\cite{wang_twc_2010} used the theoretical CDF, but used $\bar{N}$ as the number of samples to generate the CDF in their complexity figures. The same observation favoring the proposed rcK/rcKS methods holds for the magnitude-based (mag) classifiers~\cite[Sec III-A]{wang_twc_2010}.

\begin{table}
\caption{Number of Operations and Memory Usage}
\centering
\begin{tabular}{l | l l l}
\hline
Method & Multiply & Add & Memory\\
\hline\hline
Cm & $6M$ & $6M$ & $K$ \\
rcKS/rcK & $0$ & $2ML$ & $WL(K+1)$ \\
KS/K & $0$ & $2M(\log 2M + 2K)$ & $KW\bar{N}$ \\
rcKS/rcK (mag) & $2M$ & $M(L+1)$ & $WL(K+1)$ \\
KS/K (mag) & $2M$ & $M(\log M + 2K + 1)$ & $KW\bar{N}$\\
\hline
\end{tabular}
\label{table:complexity}
\end{table} 

\section{Results}
\label{sec:results}

As an example, we assume that the classification task is to distinguish between M-QAM, where $\text{M}\in\{4,16,64\}$. For comparison we also present classification result based on maximum likelihood estimation (ML).

\vspace{-0.2cm}

\subsection{Detection Performance versus SNR}
\label{sec:detection_snr}

\begin{figure}
\begin{center}
\includegraphics{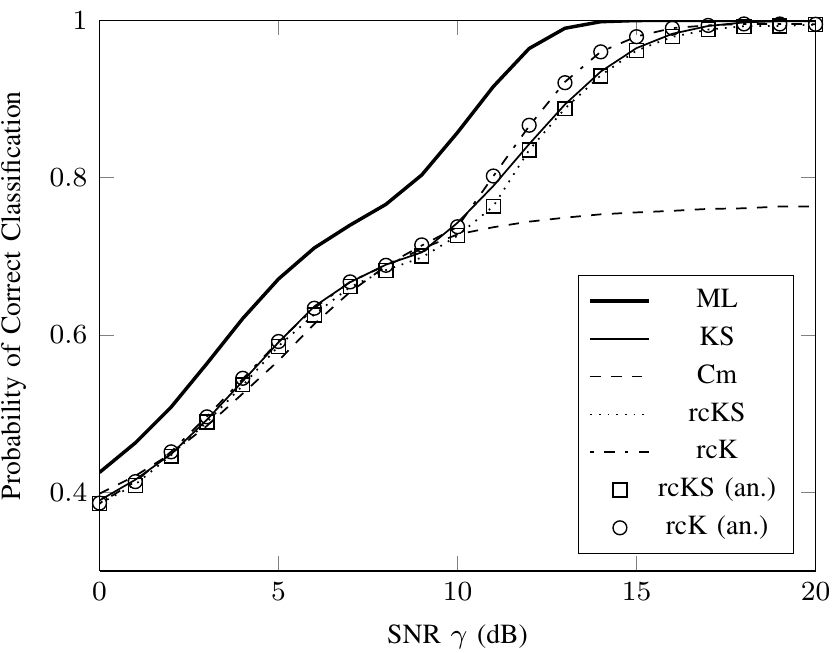}
\end{center}
\caption{Effect of varying SNR on the probability of classification with $M$=50; (an.) -- analytical result using (\ref{eq:ProbabilityOfClassification}).}
\label{fig:vary_snr}
\end{figure}

In the first set of experiments we evaluate the performance of the proposed classification method for different values of SNR. The results are presented in Fig.~\ref{fig:vary_snr}. We assume fixed sample size of $M=50$, in contrast to~\cite[Fig. 1]{wang_twc_2010} to evaluate classification accuracy for a smaller sample size. We confirm that even for small sample size, as shown in~\cite[Fig. 1]{wang_twc_2010}, Cm has unsatisfying classification accuracy at high SNR. In (10,17)\,dB region rcK clearly outperforms all detection techniques, while as SNR exceeds $\approx$17\,dB all classification methods (except Cm) converge to one. In low SNR region, (0,10)\,dB, KS, rcKS, rcK perform equally well, with Cm having comparable performance. The same observation holds for larger sample sizes, not shown here due to space constraints. Note that the analytical performance metric developed in Section~\ref{sec:analysis} for rcK and rcKS matches perfectly with the simulations. For the remaining results, we set $\gamma=12$\,dB, unless otherwise stated.

\vspace{-0.2cm}

\subsection{Detection Performance versus Sample Size}
\label{sec:detection_samples}

\begin{figure}
\begin{center}
\includegraphics{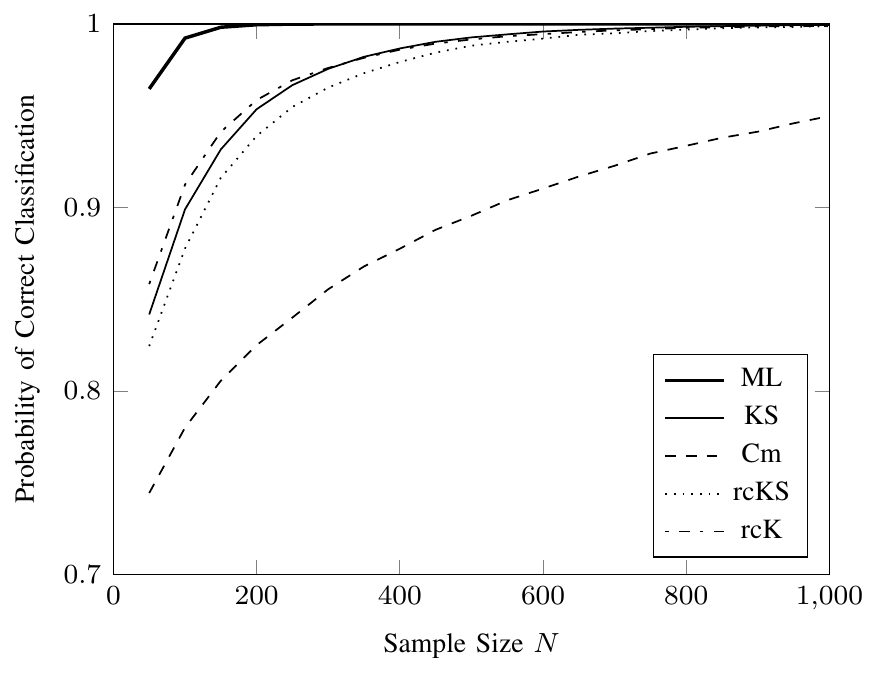}
\end{center}
\caption{Effect of varying sample size on the probability of classification with $\gamma=12$\,dB.}
\label{fig:vary_n}
\end{figure}

In the second set of experiments, we evaluate the performance of the proposed classification method as a function of sample size $M$. The result is presented in Fig.~\ref{fig:vary_n}. As observed in Fig.~\ref{fig:vary_snr}, also here Cm has the worst classification accuracy, e.g. 5\% below upper bound at $M=1000$. The rcK method performs best at small sample sizes, $50\leq M\leq300$. With $M>300$, the accuracy of rcK and KS is equal. Classification based on rcKS method consistently falls slightly below rcK and KS methods. In general, rcKS, rcK and KS converge to one at the same rate.

\vspace{-0.2cm}

\subsection{Detection Performance vs SNR Mismatch and Phase Jitter}
\label{sec:detection_snr_mismatch}

\begin{figure}
\centering
\subfigure[]{
\includegraphics{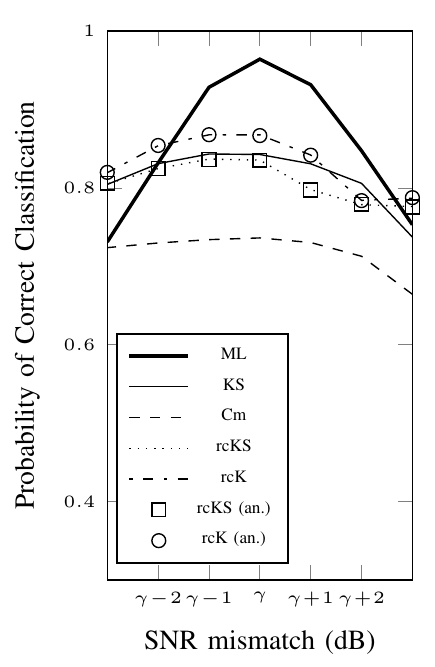}
\label{fig:vary_gamma}
}
\subfigure[]{
\includegraphics{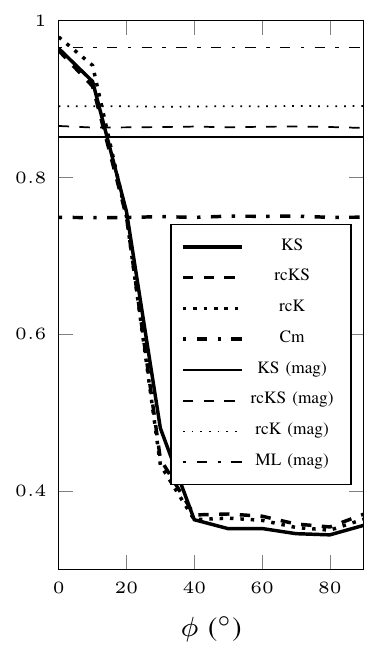}
\label{fig:vary_phase}
}
\caption{(a) Effect of SNR mismatch, nominal/true SNR=12dB; (b) effect of phase jitter, nominal SNR=15dB.; (an.) -- analytical result using (\ref{eq:ProbabilityOfClassification}), (mag) -- magnitude.}
\label{fig:mismatch}
\end{figure}

In the third set of experiments we evaluate the performance of the proposed classification method as a function of SNR mismatch and phase jitter. The result is presented in Fig.~\ref{fig:mismatch}. In case of SNR mismatch, Fig.~\ref{fig:vary_gamma}, our results show the same trends as in~\cite[Fig. 4]{wang_twc_2010}; that is, all classification methods are relatively immune to SNR mismatch, i.e. the difference between actual and maximum SNR mismatch is less than 10\% in the considered range of SNR values. This justifies the selection of the limited set of SNR values $W$ for complexity evaluation used in Section~\ref{sec:complexity}. As expected, ML shows very high sensitivity to SNR mismatch. Note again the perfect match of analytical result presented in Section~\ref{sec:analysis} with the simulations. 

In the case of phase jitter caused by imperfect downconversion, we present results in Fig.~\ref{fig:vary_phase} for $\gamma=15$\,dB as in~\cite{swami_tcom_2002}, in contrast to $\gamma=12$\,dB used earlier, for comparison purposes. We observe that our method using the magnitude feature, rcK/rcKS (mag), as well as the Cm method, are invariant to phase jitter. rcK and rcKS perform almost equally well, while Cm is worse than the other three methods by $\approx$10\%. As expected, the ML performs better than all other methods. Quadrature-based classifiers, as expected, are highly sensitive to phase jitter. Note that in the small phase jitter, $\phi<10^{\circ}$, quadrature-based classifiers perform better than others, since the sample size is twice as large as in the former case.

\section{Conclusion}
\label{sec:conclusions}

In this letter we presented a novel, computationally efficient method for modulation level classification based on distribution distance functions. Specifically, we proposed to use a metric based on Kolmogorov-Smirnov and Kuiper distances which exploits the distance properties between CDFs corresponding to different modulation levels.  The proposed method results in faster MLC than the cumulant-based method, by reducing the number of samples needed. It also results in lower computational complexity than the KS-GoF method, by eliminating the need for a sorting operation and using only a limited set of test points, instead of the entire CDF.

\bibliographystyle{IEEEtran}

\end{document}